\documentstyle[preprint,aps,version2]{revtex}
\tightenlines
\begin{document}
\draft
\begin{title}
\bf{ OVERLAP, DISORDER AND DIRECTED POLYMERS: \\
A RENORMALIZATION GROUP APPROACH }
\end{title}
\author{Sutapa Mukherji\cite{eml}}
\begin{instit}
Institute of Physics, Bhubaneswar 751 005,
India
\end{instit}
\begin{abstract}
The overlap of a $d+1$ dimensional directed polymer of length
$t$ in a random medium is studied using a Renormalization Group
approach. In $d>2$ it vanishes at $T_c$ for
$t\rightarrow \infty$ as
$t^{\Sigma}$ where $\Sigma=[\frac{d-1}{3-2d}]\frac{d}{z}$ and
$z$ is the transverse spatial rescaling exponent. The same
formula holds in $d=1$ for any finite temperature and it agrees
with previous numerical simulations at $d=1$. Among other
results we also determine the scaling exponent for mutual
repulsion of two chains in the random medium.

\end{abstract}
\narrowtext

One commonly studied quantity in random systems is
the overlap of
appropriate physical variables in different states as, for
example, the overlap of magnetizations in two states for a
 spin glass system
\cite{mpv,par1}.  The distribution
function for the overlap $q$, $p(q)$, which, e.g,
in a pure Ising
system  has two $\delta$
function peaks for the two possible low temperature
ferromagnetic
states, has a nontrivial structure in the spin
glass phase because of the broken ergodicity.
Such nontriviality of the overlap distribution is the main
characteristic for spin glass type systems.
 Notwithstanding the importance of overlaps, very little is
understood analytically about it.

  Of late, a directed polymer in
a random environment is taken to be the
paradigm of disordered
systems. This is because of the strong analogy
in behavior with other
random systems and spin glass \cite{mez}, and more
so because of the
availability of exact results through Bethe
ansatz \cite{kard}, use
of various analytical and numerical methods like, nonlinear
differntial equation \cite{kpz}, transfer
 matrix \cite{kard1,kim,der1}
 and other approaches.
 Even in this situation, the question of overlap
 remains a mind
boggling issue \cite{kim,der1}. In this paper,
we implement a renormalization
group (RG) approach to obtain the scaling
behavior of the overlap.
To our knowledge, no such analytical result for overlap is
known for any other realistic random system.

The $d+1$
dimensional directed polymers are random walks stretched
along a particular direction with
fluctuations in the transverse
$d$ dimensional space. These string like objects appear
in various random systems of interface
 fluctuations and pinning
\cite{kard},
crystal growth \cite{kpz}, spin glasses \cite{dersp} etc.
 Most
of the recent work done on this
problem attempt to understand the low
temperature (strong disorder)
phase which, as a matter of fact,
 is the only possible phase in $1+1$ dimensions.
The problem in $1+1$ dimensions is almost settled with
relevant exponents known exactly \cite{huhenf}.
The two important exponents are $\chi$ and
$\zeta=1/z$, which describe the free energy fluctuation and
transverse size as the length $t\rightarrow \infty$, namely
$f\sim t^{\chi/z}$ and $<x>\sim t^{\zeta}$. For $d=1$,
$\chi=1/2$ and $z=3/2$, with $\chi+z=2$.
 However, precise values of these
exponents in higher dimensions and their exact dimensional
dependences are not yet well understood mainly because of the
lack of any perturbative fixed point \cite{kost1}.
For high enough
dimensions ($d>2$) it is found that there
is a phase transition
\cite{imb,code} from the
 high temperature (weak disorder) to the low
temperature phase (strong disorder). What
happens at $d=2$ is not clear \cite{der1,kim},
though the consensus
seems to be against a phase transition \cite{kpz,kim}.
We mention, in passing, that several exact results are known,
especially in connection with such disorder induced phase
transitions, if the randomness is in the interaction instead of
the medium \cite{smb1}.

The high temperature phase of a directed polymer in a random
potential  in $d> 2$
 is simple since there the
quenched and annealed free energies are equal.
The situation is more complex in the low temperature
phase because of the nonzero overlap and the
subsequent nonanalyticity of the free energy that
supports the
hypothesis of the coexistance of several pure
states \cite{dersp,mez} - a
phenomena reminiscent of spin glasses.  Partial
information regarding the
thermal, geometrical properties and a few
related distribution
functions in the low temperature phase are available from
Monte Carlo simulations, expansion methods and
transfer matrix
techniques \cite{code,der1}.  However, at present the overlap,
especially near the critical point, seems to elude these
techniques.
 In this paper our main focus is at this transition
temperature. Unlike the tree problem \cite{deri}
which, in some
way, corresponds to the mean field limit $d\rightarrow \infty$,
our result is true for finite dimensions. [See, e.g., Ref.
\cite{hf} for the peculiarities of the Cayley tree problem.]

 Similar to the concept of the overlap of two different
magnetization states \cite{mpv},
here in the DP picture, the overlap means the
number of common spatial points  visited by two different
configurations of the polymer. Introducing two different
configurations of a polymer is equivalent to starting with
the original polymer  with a replica.
 The overlap is then the average number of contacts
of these two polymers (see below for a more
precise definition).
The procedure for the evaluation of the overlap would be to
 introduce a new interaction that penalizes such
contacts with coupling constant $v_0$.
The overlap follows, as statistical mechanics
prescribes, from the calculation of the appropriate derivative
of the free
energy of such
an interacting system.

In the path integral
representation the working hamiltonian for two
 interacting directed
polymers
\begin{eqnarray}
{\cal H}=\int\limits_{0}^{t} d\tau \
\ \big[\sum_{i=1,2}\left(\frac{\gamma}{2}
{{\bf\dot{x_i}}}^2(\tau)+\frac{\lambda}{2\gamma}
V({\bf x_i}(\tau),\tau)\right)+
\frac{\lambda}{2\gamma}v_0\ \delta ({\bf x}_{12}(\tau))\big]
 \label{eq:two}
\end{eqnarray}
where ${\bf x}_i(t)$ is the $d$ dimensional
spatial coordinate of the
$i$th chain at the contour length $t$,
$ \dot{{\bf x}_i}(t)=\frac{d{\bf x}_i(t)}{dt}$,
and ${\bf x}_{12}$ is the
relative seperation of the two chains.
The  first two terms represent the  entropic
fluctuations of two free Gaussian chains with $\gamma$
as the bare line tension. $V$ corresponds to
the space and time dependent  random
potential seen by the two chains interacting at same $t$ with
$\delta$ function potential of
strength $\lambda v_0/(2\gamma)$.
The significance of the peculiar
factor $\frac{\lambda}{2\gamma}$
with the random potential and  with the coupling constant
$v_0$ will be clear from discussions later.
The  random potential is taken to be
uncorrelated, normally distributed \cite{kpz} with
\begin{equation}
\overline{V({\bf x},\tau)V({\bf y},\tau ')}=
2 \Delta\ \delta({\bf
x} -{\bf y})\delta(\tau-\tau')
\label{eq:three}
\end{equation}
where the overbar stands for the averaging over the disorder.
 The disorder is quenched so that the average of
$\ln W$, where $W$ is the appropriate partition function,
is needed.

 The overlap for the above system in the
continuum limit  can be
 precisely defined as
\begin{equation}
q(t)=\frac{1}{t}\int_0^t
d\tau\ \delta({\bf{x}}_{12}(\tau)). \label{eq:one}
\end{equation}
It can be obtained from
 the relation
$q(t)=- \frac{1}{t}\frac{df_2(v_0,t)}{{dv_0}}\mid_{v_0=0}$
where
 $f_2(v_0,t)$ is the
free energy for the Hamiltonian in Eq.\ref{eq:two}.
A scaling form
\begin{equation}
f_2(v_0,t)=
{ t}^{\chi/z}{ f}(v_0\ {t}^{-\phi/z})\label{eq:fr}
\end{equation}
is expected with $\phi$ determining the crossover exponent.
 This implies $q=t^{\Sigma}Q(v_0t^{-\phi/z})$ where
\begin{equation}
\Sigma=(\chi-\phi-z)/z
\end{equation}
This particular problem
in a discrete version at $d=1$ has been studied
by Mezard numerically in
 \cite{mez}.
 His simulation results
are consistent with $\zeta=1/z=2/3$, $\chi=1/2$,
and $\phi/z=-2/3$.
A corollary is that the behavior of one chain
 remains unaffected
by the presence of the other. This need not be surprising
because the ``screening'' type effects in ordinary
self and mutually
avoiding polymers are finite density phenomena.
Since at $d=1$
$\chi-\phi-z=0$, one obtains $q(v_0,t)\sim
q(v_0 t^{-\phi/z})$.
One of our aims is to determine $\phi$.

Here we use the continuum formulation and map the problem
 to a  KPZ type nonlinear
differential equation for the free energy \cite{kpz,med}.  A
dynamic renormalization group approach, {\it \`{a} la}
Ref.\cite{med},  following a perturbative
calculation in a Fourier conjugate space, is
 developed to study
this equation. In the process, we obtain the scaling exponent
for the interaction and establish that in any arbitrary
dimensions $\chi$ and $z$ remain the same as
those of the single
chain problem, as one would expect, even in the
presence of the
interaction. A series for the renormalized
coupling constant can
be identified by collecting the appropriate terms from the
perturbative series.  The recursion relation for the coupling
constant, found after the use of the momentum
shell technique,
manifests the scaling form of the mutual interaction and the
overlap. The exponents obtained through this
 process are for the
critical point $T_c$ for $d>2$ but for the finite temperature
phase for $d=1$. An appeal to finite size scaling
 then enables us
to extend the result to $T\neq T_c$ for $d>2$.

We consider two chains which are tied at one end ($t=0$)
at the origin of the $d$ dimensional space and extended upto
${\bf x}_1$ and ${\bf x}_2$ at length $t$.
The partition function $W({\bf x}_1,{\bf x}_2,t)$,
 which is basically a sum of
the Boltzmann weights of all configurations of  two such
chains can be written in the path integral form as
\begin{eqnarray}
W({\bf x}_1,{\bf x}_2,t)=\int_{(0,0,0)}^{({\bf x}_1,{\bf
x}_2,t)}{\cal D}{{\bf x}'_1}\ {\cal D}{{\bf x}'_2}\
\exp{\big[-\cal H\big]},
\label{eq:seven}
\end{eqnarray}
where $\int {\cal D}{\bf x}'_1\ {\cal D}{\bf x}'_2$
stands for all
possible paths of the two polymers and ${\cal H}$ is given by
Eq. \ref{eq:two}.
This implies that the partition function satisfies
a Schrodinger type equation written suppressing the
 argument of
$W$
\begin{eqnarray}
\frac{\partial}{\partial t} W=
\left[\gamma\sum_{i=1,2}\nabla_i^2
+\frac{\lambda}{2\gamma}g_0({\bf x}_1,{\bf x}_2,t) \right]
W,\label{eq:eight}
\end{eqnarray}
where $g_0({\bf x}_1,{\bf x}_2,t)=V({\bf x}_1,t)+
V({\bf x}_2,t)+
v_0\delta({\bf x}_{12})$
appear as the potential.

Our approach starts with another version of the above
equation for the free energy
$h({\bf x}_1,{\bf x}_2,t)=({2\gamma}/\lambda)\ln\
W({\bf x}_1,{\bf
x}_2,t)$
  which satisfies, again suppressing the arguments,
\begin{eqnarray}
\frac{\partial}{\partial t} h
=\sum_{i=1,2}[\gamma\nabla_i^2 h
+\frac{\lambda}{2}(\nabla_ih)^2]+g_0.\label{eq:nine}
\end{eqnarray}
The impressive feature of this equation is that the parameter
$\lambda$, which was previously controlling
the random potential
and the mutual interaction between the chains in the original
hamiltonian in Eq.\ref{eq:two}, now appears only with the
nonlinear term. Eq.~\ref{eq:nine} can be decoupled when
there is no mutual interaction between the chains
 and such a decoupled equation can
be solved exactly when
$\lambda=0$ \cite{edwil}.  One can then make a small
perturbation in
 the nonlinearity (i.e. in $\lambda$
 which is effectively equivalent to introducing
a small disorder into the problem.  The recursion
 relation for $v_0$,
obtained after perturbation in the nonlinearity
$\lambda$, gives the influence of disorder on $v_0$.

A glance at Eq.\ref{eq:nine} shows that under the
transformation
$x\rightarrow b\ x$ and $t\rightarrow b^z\ t$
the parameters of the equation change as
\begin{eqnarray}
(\gamma,\lambda,v_0)\rightarrow (b^{z-2} \gamma,\
b^{\chi+z-2}\lambda,\ b^{z-d-\chi}\ v_0). \label{eq:ten}
\end{eqnarray}
Therefore in the absence of the nonlinearity
i.e when $\lambda=0$,
$z=z_0=2$ and $\chi=2-d$ keep $\gamma$ and $v_0$ invariant.
This scaling furthermore ensures the speciality of $d=d_c=2$
since for $d<d_c$ a small amount of nonlinearity becomes
relevant with the growth of the length scale. Such
speciality of
$d_c$ is reflected later in the perturbative series.
At this level one finds the necessity of going
beyond the simple
scaling analysis to carry out the RG analysis since the
scaling dimension  of $v_0$, $z-\chi-d$, vanishes
at $d=1$, while numerically it is found
to be $-1$ \cite{mez}.

The formal solution of Eq.\ref{eq:nine} in (${\bf K},{\bf
k},\omega$) space, Fourier conjugate to $({\bf x}_1,{\bf
x}_2,t)$, is given by
\begin{eqnarray}
&h({\bf K},{\bf k},\omega)=G_0({\bf K},{\bf k},\omega)
g_0({\bf K},{\bf k},\omega)
-{\lambda}/2\ G_0({\bf K},{\bf k},\omega)\times\nonumber \\
&\int_{p,q,\Omega}
[{\bf p} \cdot ({\bf K}-{\bf p})\ +{\bf q}\cdot
({\bf k}-{\bf q})]
h({\bf p},{\bf q},\Omega)h({\bf K}-{\bf p},
{\bf k}-{\bf q},\omega-\Omega),\label{eq:eleven}
\end{eqnarray}
where
$G_0({\bf K},{\bf k},\omega)=
(\gamma({\bf K}^2+{\bf k}^2)-i\omega)^{-1}$
 represents the bare propagator and $\int_{p,q,\Omega}=
\int \frac{d\Omega}{2\pi}\frac{d{\bf p}
d{\bf q}}{(2\pi)^{2d}}$.
The fact that the random potential and the
interaction are in the same footing in the above
equation is now
 utilised in defining the effective propagator
 $G({\bf K},{\bf k},\omega)$
and the effective coupling constant $v$ as
\begin{equation}
h({\bf K},{\bf k},\omega)=G({\bf K},{\bf k},\omega)
[V({\bf K},\omega)\delta({\bf k})+
V({\bf k},\omega)\delta({\bf K})+v
\delta({\bf K}+{\bf k})\delta(\omega)]. \label{eq:thirteen}
\end{equation}
Note that such a restriction on momenta is
imposed automatically
by the RHS of Eq.\ref{eq:eleven}.

Now we are in a position to initiate the perturbative
series, the terms of which  after disorder averaging leads to
``closed loop diagrams''. [See, e.g., Ref.\cite{med}].
 Here we shall consider terms
upto $O(\lambda^2)$ and $O(v_0)$.
 Collecting the appropriate terms from the
perturbative series for the renormalized propagator
satisfying
either ${\bf K}=0$ or ${\bf k}=0$, one
obtains a series identical to the renormalized propagator
for a single chain in a random medium \cite{kpz,med}.
 It becomes evident from
the series that there is no contribution at
$O(\lambda)$ since such terms either donot have any loop or
even if they do, they contribute to $O(v_0^2)$, a
higher order
term which we are not considering here.
For convenience, we cite
the series obtained for a single chain propagator
with momentum
variable
 ${\bf K}$
\begin{eqnarray}
& G({\bf K},{\bf k},\omega)=G_0({\bf K},{\bf k},\omega)
+ C(-\lambda/2)^2
G_0^2({\bf K},{\bf k},\omega)(2\Delta)\int_{q,\Omega}
{\bf q}
\cdot ({\bf K}-{\bf q})\ {\bf q}\cdot
{\bf K}\times \nonumber \\
& G_0({\bf K}-{\bf q},{\bf k},\omega-\Omega)G_0({\bf q},
0,\Omega)
G_0(-{\bf q},0,-\Omega) \label{eq:fourteen}
\end{eqnarray}
where $C=4$ counts all the possible ways of noise
contraction
and ${\bf k}=0$ in this case.
The  terms which
 contribute to the renormalization of the vertex
follow the constraint ${\bf K}+{\bf k}=0$. One can
write down the
series for $vG({\bf K},-{\bf K},0)$ to obtain an equation in
which the vertex  and the
propagator $G({\bf K},-{\bf K},0)$ renormalization
take place in a
combined fashion and interestingly can also be isolated.
To $O(\lambda^2)$ the  series for the renormalized
propagator
$G({\bf K},-{\bf K},0)$ is same as Eq.\ref{eq:fourteen}
with $C=8$ and ${\bf k}=-{\bf K}$ and $v$ is given
by the series
\begin{eqnarray}
v=v_0
+& 8(-\frac{\lambda}{2})^2(2v_0\Delta)\int_{q,\Omega}
({\bf q}\cdot ({\bf K}-{\bf q}))^2
G_0({\bf q},-{\bf K},\Omega)
G_0({\bf K}-{\bf q},0,-\Omega)\times\nonumber\\
& G_0({\bf q},-{\bf q},0)G_0(0,{\bf q}-{\bf K},
\Omega).\label{eq:sixteen}
\end{eqnarray}
In a digrammatic representation the second term on the RHS of
this equation would
correspond to an exchange type diagram.
At the face value,
 the series of $G({\bf K},-{\bf K},0)$ does not resemble
 the series for the single chain
propagator $G({\bf K},0,\omega)$,
but it is easy to show that there is no change in the
 renormalization of the
 line tension $\gamma$ from that of Ref.\cite{kpz}
in the long wavelength
limit.  This, furthermore, confirms that in any arbitrary
dimensions, so far as the free
energy, lateral extension are concerned,
exponents do not change
even after
the inclusion of the mutual interaction.

The second term of Eq.\ref{eq:sixteen} involves a
 momentum integration with an upper
cutoff $\Lambda$ which indicates the resolution upto
which the system is probed. To study the
variation of $v$ with
length scale, we execute an RG procedure
consisting of two steps
(i) an integration over the momentum
shell between the momenta $\Lambda$
and $\Lambda \exp{(-\delta l)}$
and (ii)  the momentum rescaling ${\bf k}\rightarrow {\bf k}
\exp{(-\delta l)}$ which restores the upper
cutoff $\Lambda$ as before.
Now after carrying out the first step and the integration
over $\Omega$ we obtain an effective coupling
 constant (in the
long wavelength limit) differing from $v_0$ by a term
$K_d{\bar{\lambda}}^2v_0\delta l/2$. Here
 $K_d=\frac{S_d}{(2\pi)^d}$ with $S_d$ as the
 surface area of the
unit $d$ dimensional sphere and
${\bar{\lambda}}^2=\frac{\lambda^2\Delta}{\gamma^3}$ is
dimensionless.
This additional term is the fluctuation contribution
of the disorder and crucial for the RG analysis.
A little manipulation after the
rescaling (step (ii)) produces
the recursion relation
 \begin{equation}
\frac{dv}{dl}=(z-\chi-d)v+\frac{{\bar{\lambda}}^2 v\
K_d}{2},\
\label{eq:eighteen}
\end{equation}
with $v=v_0$ when $l=0$.
The recursion for ${\bar{\lambda}}$ is quoted below from
Ref.\cite{kpz}
\begin{equation}
\frac{d{\bar\lambda}}{dl}=\frac{2-d}{2}{{\bar\lambda}}+
K_d\frac{(2d-3)}{4d}
{\bar\lambda}^3.\label{eq:19}
\end{equation} which has a fixed point
${{\bar\lambda}^*}=[2d(2-d)/((3-2d)K_d)]^{1/2}$.
At $d=1$ it is a stable fixed point which  shows that the
phase is influenced by the disorder  at all temperatures.
There is no physical fixed point for $1.5<d<2$.
The nontrivial fixed point becomes
unstable for $d>2$. From the
flow, one concludes that this unstable
 fixed point corresponds
to a critical point that seperates the two phases
 dominated by the entropy (high temperature phase)
or by the disorder (low temperature phase).
 The situation at
$d=2$ is more complicated since disorder is
marginally relevent
\cite{med,lihan}.
Using the relation $\chi+z=2$, which is a
consequence of Galelian
invariance, and ${\bar \lambda}= {\bar \lambda}^*$ we
find from
Eq.\ref{eq:eighteen}
\begin{equation}
\phi=(2+d-2z)+{d(d-2)}/{(3-2d)}\label{eq:exp1}.
\end{equation}
At $d=1$ exact values $z=3/2$ and $\chi=1/2$ yield the scaled
variable as $vt^{2/3}$. The
remarkable feature is that in Eq.\ref{eq:eighteen} the term
$z-\chi-d$ which
originates from the simple scaling analysis
of Eq.\ref{eq:ten},
 vanishes at $d=1$
and the entire $t$ dependence of the scaled variable comes
only from the fluctuation part.
The scaling exponent of $v$
 matches  with
the numerical prediction of Mezard at $d=1$
{}~\cite{mez}.
Now, using the exponent $\phi$, Eq.\ref{eq:exp1}
 we find
\begin{equation}
\Sigma=
d({d-1})/[z(3-2d)]
\label{eq:exp2}
\end{equation}
Going back to the original problem of a single DP,
 the overlap is
obtained by setting $v_0=0$. We therefore obtain
\begin{equation}
q\sim t^{\Sigma},\label{eq:e3}
\end{equation}
where, as mentioned before, this is the overlap in the low
temperature phase for $d<1.5$ but at the critical point for
$d>2$.
The exponent $\Sigma$ vanishes for $d=1$, as has
been found in
Ref.\cite{mez}.
Since $\Sigma<0$ for $d>2$, the overlap disappears at the
transition temperature when the thermodynamic limit is
approached - as one should expect.

We extend our $T=T_c$ result to the critical region
by invoking a
finite size scaling
hypothesis, \cite{bar} $q={\xi}^{\Sigma_1}Q(t/{\xi})$
where $\xi$ is the longitudinal (parallel to $t$) correlation
length and $\xi\sim \mid T-T_c\mid^{-\nu}$ near the critcal
point.  Comparing with
Eq.\ref{eq:e3}, we find $\Sigma_1=\Sigma$, so that, in the
thermodynamic limit
\begin{equation}
q\sim \mid T-T_c\mid^{-\nu\Sigma}\label{eq:e5}
\end{equation}
as $T\rightarrow T_c$.
Unfortunately the value of $\nu$ is still not known with
confidence \cite{kim,der1}.

It is possible to explain the effect of random
environment more
physically.
In the situation where the disorder dominates the
physics, the
chain is swollen to take advantage of the occasional
traps that
lower the energy. The loss in the entropy is
offset by the gain
in the energy, yielding $\zeta>1/2$. In this scenario, it is
therefore expected that the repulsion with another chain will
have no significant effect. Hence to the
leading order, $v_0$ is
not to have any effect on the renormalization
of the properties
of the polymers. However, with two chains there
will be a certain
amount of overlap in their attempts to take the
advantage of the
same traps. On a bigger scale, such closely
spaced traps would
appear as an inetraction between the chains. This
leads to the
renormalization of $v_0$ changing its scaling exponent.

In summary, we have shown through one loop RG analysis
that the behavior of a single chain
in the random medium remains unaffected even if we introduce
another chain interacting with it through a
short-range repulsive
interaction. At $d=1$ our results show
 a finite overlap at all temperatures
indicating a strong coupling phase and the exponents match
exactly with Mezard's numerical simulations.
It may not be surprising that the 1 loop result gives exact
results in $1$ dimension, because it is
known to happen for all
the exponents for the $d=1$ DP problem. This is mainly a
conseqence of Galelian invariance and
 fluctuation- dissipation
theorem. In higher dimensions ($d>2$), we
evaluated the scaling
exponent of the overlap at $T=T_c$. By using a
finite size scaling
ansatz, we thereafter predict the temperature
dependence of the
overlap as $T\rightarrow T_c$.

I thank S. M. Bhattacharjee for fruitful discussions and
comments on the manuscript. I also thank D. Dhar for several
suggestions.

\end{document}